\documentclass[
10pt,
prl,
floatfix,
aps,  
longbibliography,
twocolumn,
superscriptaddress,
amssymb,
tightenlines
]{revtex4-1}

\usepackage{amsmath}

\usepackage{times}
\usepackage{wasysym}
\usepackage{amssymb}
\usepackage{latexsym}
\usepackage[dvips]{graphicx}
\usepackage{graphicx}
\usepackage{dcolumn}
\usepackage{amsfonts}
\usepackage{bm}

\newcommand{\be}{\begin{equation}}
\newcommand{\ee}{\end{equation}}
\newcommand{\bea}{\begin{eqnarray}}
\newcommand{\eea}{\end{eqnarray}}

\newcommand{\rfig}[1]{Fig.\,\ref{#1}}

\newcommand{\rref}[1]{Ref.\,\onlinecite{#1}}

\def\ne{n_e}

\def\rxx{R_{xx}}
\def\ryy{R_{yy}}
\def\rh{R_{H}}

\def\easy{\left < 110 \right >}
\def\hard{\left < 1\bar10 \right >}
\def\x{\hat{x}}
\def\y{\hat{y}}
\def\nus{\nu^\star}
 
\def\ar{\alpha_R}

\def\ne{n_e}

\begin{document}
\title{Anomalous Nematic States in High Half-Filled Landau Levels}

\author{X. Fu$^\S$}
\affiliation{School of Physics and Astronomy, University of Minnesota, Minneapolis, Minnesota 55455, USA}
\author{Q. Shi$^\S$}
\altaffiliation[Present address: ]{Department of Physics, Columbia University, New York, NY, USA}
\affiliation{School of Physics and Astronomy, University of Minnesota, Minneapolis, Minnesota 55455, USA}
\author{M. A. Zudov}
\email[Corresponding author: ]{zudov001@umn.edu}
\affiliation{School of Physics and Astronomy, University of Minnesota, Minneapolis, Minnesota 55455, USA}
\author{G.\,C. Gardner}
\affiliation{Microsoft Quantum Lab Purdue, Purdue University, West Lafayette, Indiana 47907, USA}
\affiliation{Birck Nanotechnology Center, Purdue University, West Lafayette, Indiana 47907, USA}
\author{J.\,D. Watson}
\altaffiliation[Present address: ]{Microsoft Station-Q at Delft University of Technology, 2600 GA Delft, The Netherlands}
\affiliation{Birck Nanotechnology Center, Purdue University, West Lafayette, Indiana 47907, USA}
\affiliation{Department of Physics and Astronomy, Purdue University, West Lafayette, Indiana 47907, USA}
\author{M.\,J. Manfra}
\affiliation{Microsoft Quantum Lab Purdue, Purdue University, West Lafayette, Indiana 47907, USA}
\affiliation{Birck Nanotechnology Center, Purdue University, West Lafayette, Indiana 47907, USA}
\affiliation{Department of Physics and Astronomy, Purdue University, West Lafayette, Indiana 47907, USA}
\affiliation{School of Electrical and Computer Engineering and School of Materials Engineering, Purdue University, West Lafayette, Indiana 47907, USA}
\author{K. W. Baldwin}
\affiliation{Department of Electrical Engineering, Princeton University, Princeton, New Jersey 08544, USA}
\author{L. N. Pfeiffer}
\affiliation{Department of Electrical Engineering, Princeton University, Princeton, New Jersey 08544, USA}
\author{K. W. West}
\affiliation{Department of Electrical Engineering, Princeton University, Princeton, New Jersey 08544, USA}
\received{\today}

\begin{abstract}
It is well established that the ground states of a two-dimensional electron gas with half-filled high ($N \ge 2$) Landau levels are compressible charge-ordered states, known as quantum Hall stripe (QHS) phases.
The generic features of QHSs are a maximum (minimum) in a longitudinal resistance $\rxx$ ($\ryy$) and a non-quantized Hall resistance $\rh$. 
Here, we report on emergent minima (maxima) in $\rxx$ ($\ryy$) and plateau-like features in $\rh$ in half-filled $N \ge 3$ Landau levels. 
Remarkably, these unexpected features develop at temperatures considerably lower than the onset temperature of QHSs, suggesting a new ground state. 
\end{abstract}
\maketitle

The ground state of a two-dimensional electron gas (2DEG) at half-integer filling factors $\nu = i/2, i = 1,3,5,...$, can depend sensitively on the Landau level (LL) index $N$.
At $N=0$ ($\nu = 1/2,3/2$) it is a compressible composite fermion metal \cite{jain:1989}, whereas at $N = 1$ ($\nu = 5/2,7/2$) it is an incompressible fractional quantum Hall insulator formed by paired composite fermions \cite{willett:1987,pan:1999b}.
%It is a compressible composite fermion metal \cite{jain:1989} at $N=0$ ($\nu = 1/2,3/2$) and an  incompressible fractional quantum Hall insulator formed by paired composite fermions \cite{willett:1987,pan:1999b} at $N = 1$ ($\nu = 5/2,7/2$).
At $N = 2$ and several higher LLs ($\nu = i/2, i =9, 11, ...$), the competition between long-range repulsive and short-range attractive components of Coulomb interaction leads to compressible charge-ordered phases \cite{koulakov:1996,moessner:1996,fogler:1996}.
These phases can be viewed as unidirectional charge-density waves consisting of stripes with alternating integer $\nu$ (e.g., $\nu = 4$ and $\nu = 5$) and are commonly known as quantum Hall stripes (QHSs) \cite{note:fradkin}.
With few exceptions \cite{zhu:2002,pollanen:2015}, QHSs in a 2DEG confined to GaAs quantum wells align along $\easy$ crystal axis of GaAs. 
This symmetry breaking field remains enigmatic, despite many efforts to identify its origin \cite{fil:2000,koduvayur:2011,sodemann:2013,pollanen:2015}. 

The generic QHS features are a maxima (minima) in a longitudinal resistance $\rxx$ ($\ryy$), which develop at temperatures $T \lesssim 0.1$ K, and a non-quantized Hall resistance $\rh$ \cite{lilly:1999a,du:1999}. 
More precisely, QHSs form when partial filling factor $\nus = \nu - \lfloor \nu \rfloor$, where $\lfloor \nu \rfloor$
% = \max \{ m \in  \mathbb{Z}\,|\,m \le \nu$\} 
is an integral part of $\nu$, falls in the range of $0.4 \lesssim \nus \lesssim  0.6$.
The resistance anisotropy ratio $\ar\equiv \rxx/\ryy$ normally achieves a single maximal value $\ar \gg 1$ at $\delta \nu  \equiv \nus - 0.5 \approx 0$ and quickly drops to $\ar \approx 1$ at $\delta \nu \approx \pm 0.1$.
This drop occurs due to a \emph{monotonic} decrease (increase) of the $\rxx$ ($\ryy$) with $|\delta\nu|$.

In this Letter, we report on anomalous nematic states which are distinguished from QHSs by minima (maxima) in $\rxx$ ($\ryy$) and plateau-like features in $\rh$ in half-filled $N \ge 3$ Landau levels. 
The global maxima (minima) in the $\rxx$ ($\ryy$) occur away from half-filling, at $\delta \nu \approx \pm 0.08$, where the resistance anisotropy ratio attains its maximal value.
Remarkably, all these features emerge at temperatures considerably lower than the onset temperature of QHSs, which indicates possible transition to a new phase. 

The 2DEG in sample A (B) resides in a GaAs quantum well of width 29 nm (30 nm) surrounded by Al$_{0.24}$Ga$_{0.76}$As barriers.
%Both samples utilize Si doping in narrow GaAs doping wells surrounded by thin AlAs layers and positioned at a setback distance of 30 nm on both sides of the GaAs well hosting the 2DEG.
After a brief low-temperature illumination, samples nominally had the electron density $\ne \approx 3.0 \times 10^{11}$ cm$^{-2}$ and the mobility $\mu \gtrsim 2 \times 10^7$ cm$^2$V$^{-1}$s$^{-1}$.
Samples were $4\times 4$ mm squares \cite{note:hb} with indium contacts fabricated at the corners and the midsides. 
$\rxx$ ($\ryy$) was measured using a four-terminal, low-frequency lock-in technique, with the current sent between mid-side contacts along $\x \equiv \hard$ ($\y \equiv \easy$) direction. $\rh$ was measured concurrently with $\rxx$.

%%%%%%%%%%%%%%%%%%%%%%%%%
%fig 1
\begin{figure}[t]
%\resizebox{0.5\textwidth}{!}{
\includegraphics{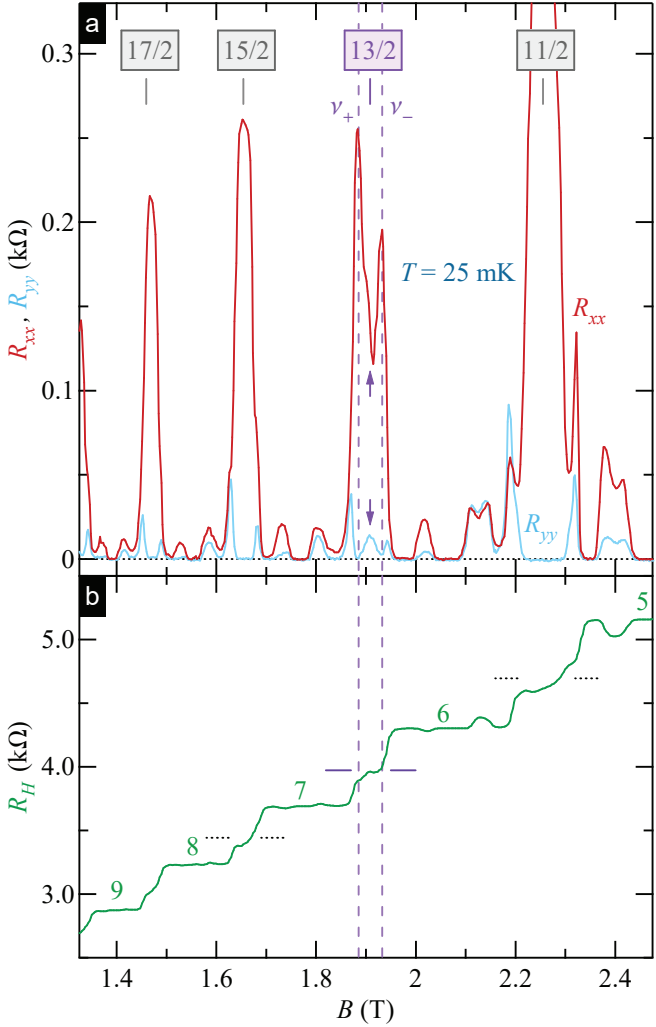}
%}
\vspace{-0.1 in}
\caption{(Color online)
(a) $\rxx$ and $\ryy$ versus $B$ measured in sample A at $T \approx 25$ mK.
% at $\bip = 0$. 
Half-integer $\nu$ are marked by $15/2, 13/2$, and $11/2$.
The $\rxx$ minimum and the $\ryy$ maximum at $\nu \approx 13/2$ are marked by $\uparrow$ and $\downarrow$, respectively.
Dashed vertical lines are drawn at $\nu_\pm = 6.5 \pm 0.08$.
(b) Hall resistance $\rh$ versus $B$.  
Solid horizontal lines, drawn at $2R_{\rm K}/13$, mark a plateau-like feature near $\nu =13/2$, while dashed horizontal lines are drawn at $2R_{\rm K}/11$ ($\nu = 11/2$) and $2R_{\rm K}/15$ ($\nu = 15/2$), where $R_{\rm K} \equiv h/e^2 = 25812.80745$ $\Omega$ is the von Klitzing constant. 
}
\vspace{-0.15 in}
\label{fig1}
\end{figure}
%%%%%%%%%%%%%%%%%%%%%%%%%

%Rxx and Ryy
In \rfig{fig1}(a) we present $\rxx$ and $\ryy$ versus magnetic field $B$ measured in sample A at $T \approx 25$ mK.  
Near $\nu = 11/2, 15/2$, and $\nu = 17/2$, $\rxx$ ($\ryy$) exhibits maxima (minima), with $\rxx \gg \ryy$, as expected of the usual QHS phases. 
Remarkably, the behavior in the vicinity of $\nu = 13/2$ is qualitatively different; even though $\rxx \gg \ryy$ (like at other $\nu=i/2$), $\rxx$ exhibits a pronounced \emph{minimum} whereas $\ryy$ shows a \emph{maximum} near half-filling.
The global maxima (minima) in $\rxx$ ($\ryy$) occur away from half-filling, namely near $\nu = 13/2 \pm 0.08$, as illustrated by vertical dashed lines.
As a result, $\ar$ becomes a \emph{non-monotonic} function of $|\delta \nu| = |\nus - 0.5|$; it is relatively small at $\delta \nu = 0$ and exhibits maxima at $\delta \nu \approx \pm 0.08$.
The variation of $\ar$ with $\nus$ is quite significant, it drops from $\ar > 600$ at $\nu  = \nu_+ \approx 6.58$ to $\ar < 10$ near half-filling.

%R_H
In \rfig{fig1}(b) we show the Hall resistance $\rh$ as a function of $B$.
Concurrent with the unexpected extrema in $\rxx$ and $\ryy$ at $\nu  = 13/2$, the Hall resistance shows a plateau-like feature, marked by solid horizontal lines drawn at $2R_{\rm K}/13$, where $R_{\rm K} \equiv h/e^2$ is the von Klitzing constant.
While the plateau-like feature in $\rh$ near $\nu = 13/2$ is very close to $2R_{\rm K}/13$, as one would expect for a developing even-denominator quantum Hall state, its appearance might be coincidental.
Indeed, steps in $\rh$ are also present near $\nu = 11/2$ and $\nu = 15/2$, albeit in these cases $\rh$ is noticeably lower than half-integer-quantized values (cf. dashed horizontal line segments drawn at $2R_{\rm K}/11$ and $2R_{\rm K}/15$.
In addition, even at $\nu=13/2$, $\rh$ often differs from $2R_{\rm K}/13$ when measured concurrently with $\ryy$.
We notice, however, that signatures of even-denominator quantum Hall states were recently observed in the $N = 3$ LL of graphene \cite{kim:2019}.
It was also established that in AlAs quantum wells, Hall quantization at $\nu = 3/2$ can occur in anisotropic setting and be accompanied by a maximum in easy resistance \cite{hossain:2019}.
Finally, fractional quantum Hall nematic states have been reported at $\nu = 7/3$ \cite{xia:2011} and $\nu =5/2$ \cite{liu:2013b} in tilted magnetic fields.

%T-dependence in Sample A
%%%%%%%%%%%%%%%%%%%%%%%%%
%fig 1
\begin{figure}[t]
%\resizebox{0.5\textwidth}{!}{
\includegraphics{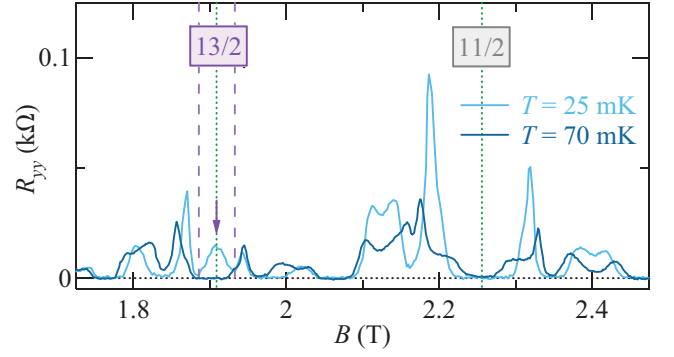}
%}
\vspace{-0.1 in}
\caption{(Color online)
$\ryy$ versus $B$ measured in the sample A at $T \approx 25$ mK (light line) and at $T \approx 70$ mK (dark line).
Half-integer $\nu$ are marked by $13/2$, and $11/2$.
}
\vspace{-0.15 in}
\label{fig2}
\end{figure}
%%%%%%%%%%%%%%%%%%%%%%%%%

The anomalous nematic state near $\nu = 13/2$ depicted in \rfig{fig1} is best observed at low temperatures.
As a glimpse at the temperature dependence, we present in \rfig{fig2} the easy resistance $\ryy$ as a function of $B$ measured in sample A at two different temperatures.
Remarkably, as the temperature is raised from  $T \approx 25$ mK to $T \approx 70$ mK, the two $\ryy$ minima near $\nu = 13/2 \pm 0.08$ and the maximum near $\nu = 13/2$ are replaced by {\em single} minimum, centered at $\nu = 13/2$ with $\ryy \approx 0$.
Such a broad minimum is a characteristic feature of the well-developed QHS phase.
In contrast, the broad minimum near $\nu = 11/2$ observed at $T \approx 25$ mK becomes narrower at $T \approx 70$ mK, consistent with previous studies of QHSs.
These data demonstrate that unexpected extrema near $\nu = 13/2$ emerge at temperatures lower than the onset temperature of QHSs.

%role of homogeneity, cooldown procedure, etc.

\begin{figure}[t]
%\resizebox{0.5\textwidth}{!}{
\includegraphics{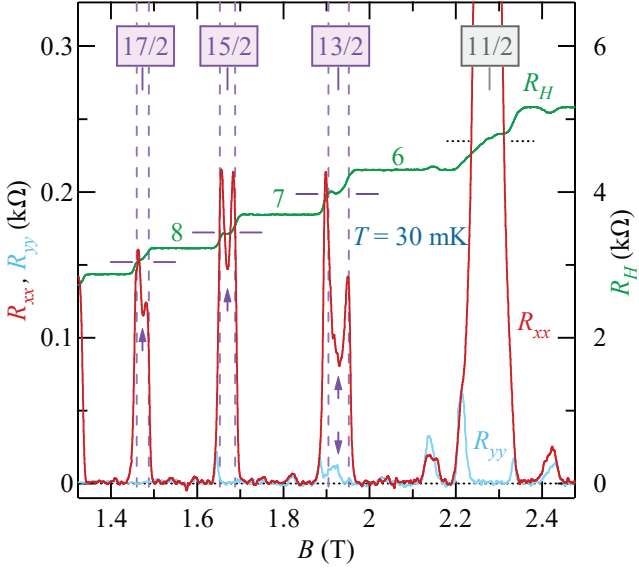}
%}
\vspace{-0.1 in}
\caption{(Color online)
$\rxx$, $\ryy$ (left axis), and $\rh$ (right axis) versus $B$ measured in sample B at $T \approx 30$ mK.
% at $\bip = 0$. 
Half-integer $\nu$ are marked by $17/2, 15/2, 13/2$, and $11/2$.
The $\rxx$ minima at $\nu  = 13/2, 15/2, 17/2$ and  the $\ryy$ maximum near $\nu = 13/2$ are marked by $\uparrow$ and $\downarrow$, respectively.
Dashed vertical lines are drawn at $\nu_\pm = i/2 \pm 0.08$, $i = 13,15,17$.
Near $\nu = i/2$ ($i = 13,15,17$), $\rh$ shows plateau-like features with $\rh \approx 2R_{\rm K}/i$, marked by solid horizontal lines. 
}
\vspace{-0.15 in}
\label{fig3}
\end{figure}

Some of our samples revealed the unexpected $\rxx$ minima not only near $\nu = 13/2$, as in \rfig{fig1}, but also near other half-integer $\nu$ \cite{note:f}.
In \rfig{fig3} we show the data obtained from sample B which exhibit pronounced $\rxx$ minima at $\nu = 13/2, 15/2$, and $17/2$.
All of these minima are accompanied by plateau-like features in $\rh$, see right axis, which assumes the values close to $2R_{\rm K}/i$, with $i = 13,15,17$, as indicated by horizontal line segments in \rfig{fig3}.
Moreover, $\rxx$ maxima occur nearly precisely at the same $\nus$ as in \rfig{fig1}, i.e., at $\nus = 1/2 \pm 0.08$, as illustrated by vertical dashed lines.
Whether or not the value of $|\delta \nu| = 0.08$ is universal remains an open question.

%%%%%%%%%%%%%%%%%%%%%%%%%
%fig 1
%%%%%%%%%%%%%%%%%%%%%%%%%

%\section{temperature dependence in sample B}
%%%%%%%%%%%%%%%%%%%%%%%%%
\begin{figure}[t]
%\resizebox{0.5\textwidth}{!}{
\includegraphics{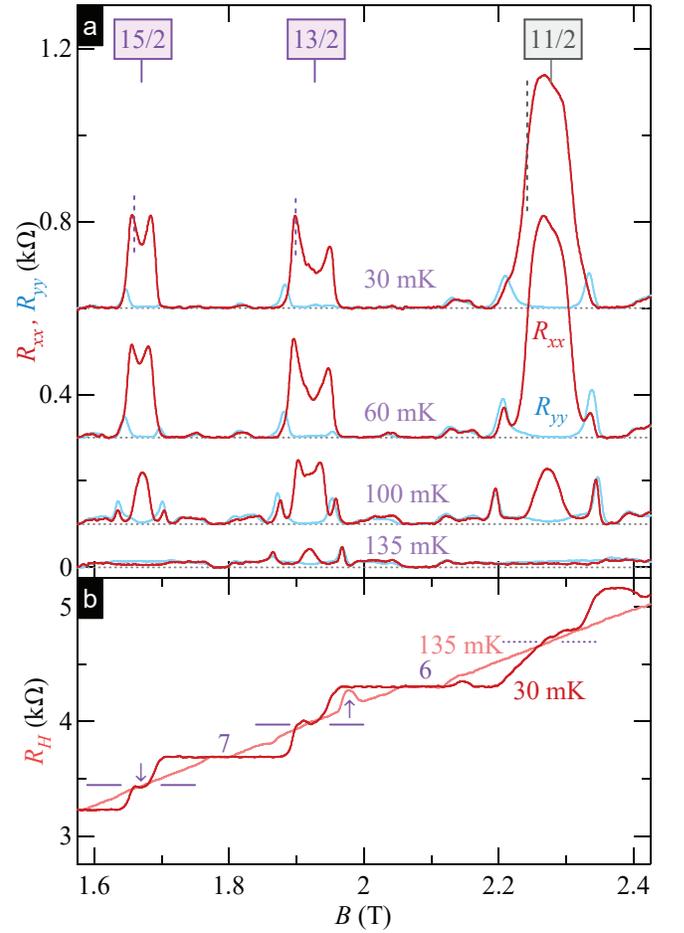}
%}
\vspace{-0.1 in}
\caption{
(Color online) 
(a) $\rxx$ (dark line), $\ryy$ (light line) versus $B$ measured in sample B at $T \approx 135$ mK (bottom), $T\approx 100$ mK (offset by 0.1 k$\Omega$), $T\approx 60$ mK (offset by 0.3 k$\Omega$), and $T \approx 30$ mK (offset by 0.6 k$\Omega$). 
Vertical dashed lines mark $\nus = 0.58.$
(b) $\rh$ versus $B$ at $T \approx 30$ mK (dark line) and at $T \approx 135$ mK (light line).
Solid horizontal lines next to the $\rh$ mark concurrent plateau-like features at $2R_{\rm K}/13$ and $2R_{\rm K}/15$, while dashed horizontal lines are drawn at $2R_{\rm K}/11$. 
}
\vspace{-0.15 in}
\label{fig4}
\end{figure}
%%%%%%%%%%%%%%%%%%%%%%%%%
We now turn to the temperature dependence in sample B which is illustrated in \rfig{fig4}(a) showing $\rxx$ (dark line) and $\ryy$ (light line) as a function of $B$ measured at different $T$, as marked. 
The Hall resistances $\rh$ measured at $T \approx 135$ mK (light line) and $T \approx 30$ mK (dark line) are shown in \rfig{fig4}(b). 
At $T \approx 135$ mK, $\rxx$ and $\ryy$ near $\nu = 11/2$ and $\nu = 15/2$ are featureless and $\rh$ is classical.
At $\nu \approx 13/2$, however, the anisotropy is already developed ($\ar \approx 6$) and $\rh$ shows a clear signature of a re-entrant integer quantum Hall state near $\nu \approx 6.72$ (as marked by $\uparrow$ in the figure), indicative of a bubble phase.
As anticipated, $\rxx$ ($\ryy$) exhibits a single maximum (minimum) at $\nu \approx 13/2$, i.e., the strongest anisotropy occurs close to half-filling, consistent with nearly all previous experiments \cite{note:as}. 
The fact that transport anisotropies in the lower-spin branches of a LL develop at higher temperatures (e.g., $\nu \approx 9/2, 13/2$) than in the upper-spin branches ($\nu \approx 11/2,15/2$) is well documented (see, e.g., \rref{lilly:1999a}).

Upon cooling to $T \approx 100$ mK, transport anisotropy with a maximum in $\rxx$ and a minimum in $\ryy$ also emerges at both $\nu \approx 11/2$ ($\ar \approx 20$) and at $\nu \approx 15/2$ ($\ar \approx 30$).
Near $\nu \approx 13/2$, however, even though the anisotropy becomes an order of magnitude stronger ($\ar \approx 60$), $\rxx$ now exhibits a pronounced \emph{minimum} near half-filling indicating an onset of the anomalous nematic state.
When the sample is cooled to $T \approx 60$ mK, the resistance anisotropy at $\nu \approx 11/2$ increases dramatically ($\ar > 300$), in agreement with previous studies.
Concurrently, we observe that the $\rxx$ minimum at $\nu \approx 13/2$ deepens and that the resistance anisotropy is \emph{reduced} by about a factor of three compared to its value at $T \approx 100$ mK.
Remarkably, the $\rxx$ near $\nu \approx 15/2$ also develops a minimum  at this temperature.
At $T \approx 30$ mK, the magnetotransport near $\nu = 11/2$ remains qualitatively unchanged, although the anisotropy ratio becomes even higher ($\ar \approx 400$).
Near $\nu  = 13/2$, however, further development of the $\rxx$ minimum and the appearance of the $\ryy$ maximum reduce the anisotropy to $\ar \approx 10$.
While we do not observe a maximum in the $\ryy$ near $\nu = 15/2$, the $\rxx$ minimum becomes more pronounced and the anisotropy reduces to $\ar < 20$.
As previously noted, the $\rxx$ minima near $\nu = 13/2$ and $\nu = 15/2$ are accompanied by plateau-like features in $\rh$, see \rfig{fig4}(b).

It is evident that the temperature dependencies near $\nu = 13/2$ and $\nu = 15/2$ are qualitatively similar.
At temperatures immediately below the onset temperature at which the QHS anisotropies sets in, the data at both filling factors exhibit normal behavior, i.e., a broad single maximum (minimum) in the $\rxx$ ($\ryy$). 
Upon cooling down further, both filling factors demonstrate the gradual development of the ``splitting'' in the $\rxx$, around half-filling, marked by a reduction of the anisotropy ratio and by the emergence of plateau-like features in the $\rh$.
We can thus conclude that, while definitely more robust in the lower spin branch, the anomalous nematic state is also supported by the upper spin branch.

%%%%%%%%%%%%%%%%%%%%%%%%%
\begin{figure}[t]
%\resizebox{0.5\textwidth}{!}{
\includegraphics{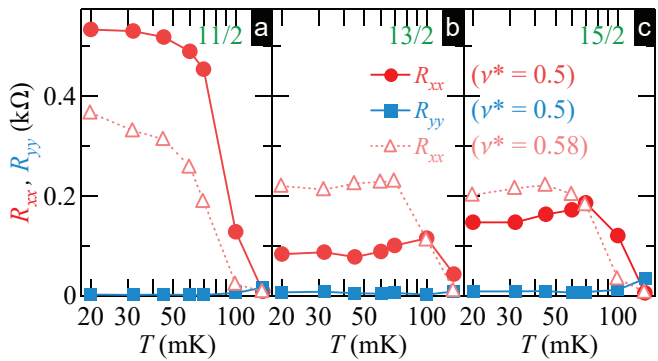}
%}
\vspace{-0.1 in}
\caption{
(Color online) 
$\rxx$ (circles), $\ryy$ (squares) versus $T$ at (a) $\nu = 11/2$, (b) $\nu = 13/2$, and (c) $\nu = 15/2$.
For comparison, $\rxx$ at $\nus = 0.58$, cf. dashed vertical lines in \rfig{fig4}(a), versus $T$ is shown by triangles.
}
\vspace{-0.15 in}
\label{fig5}
\end{figure}
%%%%%%%%%%%%%%%%%%%%%%%%%

The contrasting behavior between temperature dependencies of the $\rxx$ (circles) and of the $\ryy$ (squares) near $\nu = 13/2, 15/2$ and those near $\nu = 11/2$ are summarized in \rfig{fig5}. 
While $\rxx$ ($\ryy$) at $\nu \approx 11/2$ monotonically increases (decreases) as the temperature is lowered, $\rxx$ ($\ryy$)  at both $\nu = 13/2$ and $\nu = 15/2$ shows a clear maximum (minimum) at some intermediate ``turnover'' temperatures, $T^\star_{13/2} \approx 100$ mK and $T^\star_{15/2} \approx 70$ mK, respectively \cite{note:to}.
For comparison, we also include in \rfig{fig5} the $\rxx$ data at $\nu = i/2 + 0.08$ ($i = 11, 13, 15$), represented by triangles.
As can be seen in \rfig{fig5}(a), $\rxx$ at $\nu = 5.58$ is always smaller than that at $\nu = 11/2$ at all temperatures studied.
In contrast, $\rxx$ at $\nu =  i/2+0.08$ ($i =13,15$) is smaller than that at $\nu = i/2$ \emph{only} when $T>T_{i/2}^\star$ but not when $T<T_{i/2}^\star$.
This observation further confirms that filling factors $\nu = i/2$ ($i = 13,15$) are governed by the same physics which sets in at $T \approx T^\star_{i/2}$ and is considerably more effective at reducing the transport anisotropy at $\nu = i/2$ than away from half-filling.
Indeed, the temperature dependencies of the $\rxx$ at $\nu = 6.58,7.58$ are rather similar to that at $\nu = 5.58$.

According to the transport theory of QHS phase, which treats it as a pinned smectic \cite{macdonald:2000}, the decrease (increase) of $\rxx$ ($\ryy$) upon cooling can be attributed to the enhanced electron scattering between stripe edges.
This model, however, predicts \emph{weaker} anisotropy away from half-filling than at $\nu = i/2$, in contrast to our observations.
In addition, \rref{macdonald:2000} predicts considerably stronger $T$-dependence of $\rxx$ and $\ryy$ at $\nu = i/2$ \cite{note:mf} than away from half filling and our data do not reflect that.
Therefore, the observed dependencies on $\nu$ and $T$ are inconsistent with QHS or a nematic-to-smectic transition \cite{qian:2017}.
Instead, the observed low-temperature emergence of unexpected extrema in $\rxx$ and $\ryy$ along with the plateau-like features in $\rh$ likely reflects the formation of another competing ground state.

In addition to the temperature dependence, it is interesting to investigate the effects of the carrier density and of the in-plane magnetic field. 
Our measurements on a tunable-density Van der Pauw device with in-situ back gate have \emph{not} revealed these anomalous states at \emph{any} density from $2.2$ to $3.6\times 10^{11}$ cm$^{-2}$ \cite{shi:2017c}, as neither have those using high density [$\ne = (4.1-4.3) \times 10^{11}$ cm$^{-2}$] heterostructures \cite{fu:2019a}.
However, the carrier mobilities of samples used in the above experiments were below $1.2\times 10^{7}$ cm$^2$$V^{-1}$s$^{-1}$ and, since the anomalous nematic states form at considerably lower temperatures than QHSs, it is reasonable to expect that they are more easily destroyed by disorder. 
The absence of anomalous nematic states in these more-disordered samples yields further support to the importance of electron-electron correlations.
Measurements in tilted magnetic fields are currently under way and will be a subject of future publication.
We note, however, that the effect of in-plane magnetic field remains poorly understood even for conventional QHSs \cite{shi:2016b,shi:2016c,shi:2017c} which might complicate the interpretation of the data. 

\iffalse
Our preliminary measurements of the anomalous nematic states in tilted magnetic fields suggest that the in-plane magnetic field can easily destroys these fragile states. 
However, more experiments are needed to confirm the universality of this destruction and to understand weather or not the orientation of the in-plane field with respect to the crystal axes plays any role.
In addition, 
\fi

\begin{acknowledgments}
We thank M. Shayegan for discussions and G. Jones, S. Hannas, T. Murphy, J. Park, A. Suslov, and A. Bangura for technical support.
The work at Minnesota (Purdue) was supported by the U.S. Department of Energy, Office of Science, Basic Energy Sciences, under Award No. DE-SC0002567 (DE-SC0020138).
L.N.P. and K.W.W. of Princeton University acknowledge the Gordon and Betty Moore Foundation Grant No. GBMF 4420, and the National Science Foundation MRSEC Grant No. DMR-1420541.
A portion of this work was performed at the National High Magnetic Field Laboratory, which is supported by National Science Foundation Cooperative Agreement Nos. DMR-1157490, DMR-1644779 and the State of Florida.
\end{acknowledgments}

\small{$^\S$X.F. and Q.S. contributed equally to this work.}

%\vspace{-0.25in}
%\bibliographystyle{../../apsrev-titles}
%\bibliography{../../bibRMP1qs.1_3,footnotes}

\begin{thebibliography}{36}
\expandafter\ifx\csname natexlab\endcsname\relax\def\natexlab#1{#1}\fi
\expandafter\ifx\csname bibnamefont\endcsname\relax
  \def\bibnamefont#1{#1}\fi
\expandafter\ifx\csname bibfnamefont\endcsname\relax
  \def\bibfnamefont#1{#1}\fi
\expandafter\ifx\csname citenamefont\endcsname\relax
  \def\citenamefont#1{#1}\fi
\expandafter\ifx\csname url\endcsname\relax
  \def\url#1{\texttt{#1}}\fi
\expandafter\ifx\csname urlprefix\endcsname\relax\def\urlprefix{URL }\fi
\providecommand{\bibinfo}[2]{#2}
\providecommand{\eprint}[2][]{\url{#2}}

\bibitem[{\citenamefont{Jain}(1989)}]{jain:1989}
\bibinfo{author}{\bibfnamefont{J.~K.} \bibnamefont{Jain}},
  \emph{\bibinfo{title}{Composite-fermion approach for the fractional quantum
  Hall effect}}, \bibinfo{journal}{Phys. Rev. Lett.}
  \textbf{\bibinfo{volume}{63}}, \bibinfo{pages}{199} (\bibinfo{year}{1989}).

\bibitem[{\citenamefont{Willett et~al.}(1987)\citenamefont{Willett, Eisenstein,
  St\"ormer, Tsui, Gossard, and English}}]{willett:1987}
\bibinfo{author}{\bibfnamefont{R.}~\bibnamefont{Willett}},
  \bibinfo{author}{\bibfnamefont{J.~P.} \bibnamefont{Eisenstein}},
  \bibinfo{author}{\bibfnamefont{H.~L.} \bibnamefont{St\"ormer}},
  \bibinfo{author}{\bibfnamefont{D.~C.} \bibnamefont{Tsui}},
  \bibinfo{author}{\bibfnamefont{A.~C.} \bibnamefont{Gossard}},
  \bibnamefont{and} \bibinfo{author}{\bibfnamefont{J.~H.}
  \bibnamefont{English}}, \emph{\bibinfo{title}{Observation of an
  even-denominator quantum number in the fractional quantum Hall effect}},
  \bibinfo{journal}{Phys. Rev. Lett.} \textbf{\bibinfo{volume}{59}},
  \bibinfo{pages}{1776} (\bibinfo{year}{1987}).

\bibitem[{\citenamefont{Pan et~al.}(1999)\citenamefont{Pan, Xia, Shvarts,
  Adams, Stormer, Tsui, Pfeiffer, Baldwin, and West}}]{pan:1999b}
\bibinfo{author}{\bibfnamefont{W.}~\bibnamefont{Pan}},
  \bibinfo{author}{\bibfnamefont{J.-S.} \bibnamefont{Xia}},
  \bibinfo{author}{\bibfnamefont{V.}~\bibnamefont{Shvarts}},
  \bibinfo{author}{\bibfnamefont{D.~E.} \bibnamefont{Adams}},
  \bibinfo{author}{\bibfnamefont{H.~L.} \bibnamefont{Stormer}},
  \bibinfo{author}{\bibfnamefont{D.~C.} \bibnamefont{Tsui}},
  \bibinfo{author}{\bibfnamefont{L.~N.} \bibnamefont{Pfeiffer}},
  \bibinfo{author}{\bibfnamefont{K.~W.} \bibnamefont{Baldwin}},
  \bibnamefont{and} \bibinfo{author}{\bibfnamefont{K.~W.} \bibnamefont{West}},
  \emph{\bibinfo{title}{Exact Quantization of the Even-Denominator Fractional
  Quantum Hall State at $\nu = 5/2$ Landau Level Filling Factor}},
  \bibinfo{journal}{Phys. Rev. Lett.} \textbf{\bibinfo{volume}{83}},
  \bibinfo{pages}{3530} (\bibinfo{year}{1999}).

\bibitem[{\citenamefont{Koulakov et~al.}(1996)\citenamefont{Koulakov, Fogler,
  and Shklovskii}}]{koulakov:1996}
\bibinfo{author}{\bibfnamefont{A.~A.} \bibnamefont{Koulakov}},
  \bibinfo{author}{\bibfnamefont{M.~M.} \bibnamefont{Fogler}},
  \bibnamefont{and} \bibinfo{author}{\bibfnamefont{B.~I.}
  \bibnamefont{Shklovskii}}, \emph{\bibinfo{title}{Charge density wave in
  two-dimensional electron liquid in weak magnetic field}},
  \bibinfo{journal}{Phys. Rev. Lett.} \textbf{\bibinfo{volume}{76}},
  \bibinfo{pages}{499} (\bibinfo{year}{1996}).

\bibitem[{\citenamefont{Moessner and Chalker}(1996)}]{moessner:1996}
\bibinfo{author}{\bibfnamefont{R.}~\bibnamefont{Moessner}} \bibnamefont{and}
  \bibinfo{author}{\bibfnamefont{J.~T.} \bibnamefont{Chalker}},
  \emph{\bibinfo{title}{Exact results for interacting electrons in high Landau
  levels}}, \bibinfo{journal}{Phys. Rev. B} \textbf{\bibinfo{volume}{54}},
  \bibinfo{pages}{5006} (\bibinfo{year}{1996}).

\bibitem[{\citenamefont{Fogler et~al.}(1996)\citenamefont{Fogler, Koulakov, and
  Shklovskii}}]{fogler:1996}
\bibinfo{author}{\bibfnamefont{M.~M.} \bibnamefont{Fogler}},
  \bibinfo{author}{\bibfnamefont{A.~A.} \bibnamefont{Koulakov}},
  \bibnamefont{and} \bibinfo{author}{\bibfnamefont{B.~I.}
  \bibnamefont{Shklovskii}}, \emph{\bibinfo{title}{Ground state of a
  two-dimensional electron liquid in a weak magnetic field}},
  \bibinfo{journal}{Phys. Rev. B} \textbf{\bibinfo{volume}{54}},
  \bibinfo{pages}{1853} (\bibinfo{year}{1996}).

\bibitem[{not({\natexlab{a}})}]{note:fradkin}
\bibinfo{note}{With further consideration of thermal and quantum fluctuations,
  several electron liquid crystal-like phases have also been proposed
  \citep{fradkin:1999}.}

\bibitem[{\citenamefont{Zhu et~al.}(2002)\citenamefont{Zhu, Pan, Stormer,
  Pfeiffer, and West}}]{zhu:2002}
\bibinfo{author}{\bibfnamefont{J.}~\bibnamefont{Zhu}},
  \bibinfo{author}{\bibfnamefont{W.}~\bibnamefont{Pan}},
  \bibinfo{author}{\bibfnamefont{H.~L.} \bibnamefont{Stormer}},
  \bibinfo{author}{\bibfnamefont{L.~N.} \bibnamefont{Pfeiffer}},
  \bibnamefont{and} \bibinfo{author}{\bibfnamefont{K.~W.} \bibnamefont{West}},
  \emph{\bibinfo{title}{Density-Induced Interchange of Anisotropy Axes at
  Half-Filled High Landau Levels}}, \bibinfo{journal}{Phys. Rev. Lett.}
  \textbf{\bibinfo{volume}{88}}, \bibinfo{pages}{116803}
  (\bibinfo{year}{2002}).

\bibitem[{\citenamefont{Pollanen et~al.}(2015)\citenamefont{Pollanen, Cooper,
  Brandsen, Eisenstein, Pfeiffer, and West}}]{pollanen:2015}
\bibinfo{author}{\bibfnamefont{J.}~\bibnamefont{Pollanen}},
  \bibinfo{author}{\bibfnamefont{K.~B.} \bibnamefont{Cooper}},
  \bibinfo{author}{\bibfnamefont{S.}~\bibnamefont{Brandsen}},
  \bibinfo{author}{\bibfnamefont{J.~P.} \bibnamefont{Eisenstein}},
  \bibinfo{author}{\bibfnamefont{L.~N.} \bibnamefont{Pfeiffer}},
  \bibnamefont{and} \bibinfo{author}{\bibfnamefont{K.~W.} \bibnamefont{West}},
  \emph{\bibinfo{title}{Heterostructure symmetry and the orientation of the
  quantum Hall nematic phases}}, \bibinfo{journal}{Phys. Rev. B}
  \textbf{\bibinfo{volume}{92}}, \bibinfo{pages}{115410}
  (\bibinfo{year}{2015}).

\bibitem[{\citenamefont{Fil}(2000)}]{fil:2000}
\bibinfo{author}{\bibfnamefont{D.~V.} \bibnamefont{Fil}},
  \emph{\bibinfo{title}{Piezoelectric mechanism for the orientation of stripe
  structures in two-dimensional electron systems}}, \bibinfo{journal}{Low Temp.
  Phys.} \textbf{\bibinfo{volume}{26}}, \bibinfo{pages}{581}
  (\bibinfo{year}{2000}).

\bibitem[{\citenamefont{Koduvayur et~al.}(2011)\citenamefont{Koduvayur,
  Lyanda-Geller, Khlebnikov, Cs\'athy, Manfra, Pfeiffer, West, and
  Rokhinson}}]{koduvayur:2011}
\bibinfo{author}{\bibfnamefont{S.~P.} \bibnamefont{Koduvayur}},
  \bibinfo{author}{\bibfnamefont{Y.}~\bibnamefont{Lyanda-Geller}},
  \bibinfo{author}{\bibfnamefont{S.}~\bibnamefont{Khlebnikov}},
  \bibinfo{author}{\bibfnamefont{G.}~\bibnamefont{Cs\'athy}},
  \bibinfo{author}{\bibfnamefont{M.~J.} \bibnamefont{Manfra}},
  \bibinfo{author}{\bibfnamefont{L.~N.} \bibnamefont{Pfeiffer}},
  \bibinfo{author}{\bibfnamefont{K.~W.} \bibnamefont{West}}, \bibnamefont{and}
  \bibinfo{author}{\bibfnamefont{L.~P.} \bibnamefont{Rokhinson}},
  \emph{\bibinfo{title}{Effect of Strain on Stripe Phases in the Quantum Hall
  Regime}}, \bibinfo{journal}{Phys. Rev. Lett.} \textbf{\bibinfo{volume}{106}},
  \bibinfo{pages}{016804} (\bibinfo{year}{2011}).

\bibitem[{\citenamefont{{Sodemann} and {MacDonald}}(2013)}]{sodemann:2013}
\bibinfo{author}{\bibfnamefont{I.}~\bibnamefont{{Sodemann}}} \bibnamefont{and}
  \bibinfo{author}{\bibfnamefont{A.~H.} \bibnamefont{{MacDonald}}},
  \emph{\bibinfo{title}{Theory of Native Orientational Pinning in Quantum Hall
  Nematics}}, \bibinfo{journal}{arXiv:1307.5489}  (\bibinfo{year}{2013}).

\bibitem[{\citenamefont{Lilly et~al.}(1999)\citenamefont{Lilly, Cooper,
  Eisenstein, Pfeiffer, and West}}]{lilly:1999a}
\bibinfo{author}{\bibfnamefont{M.~P.} \bibnamefont{Lilly}},
  \bibinfo{author}{\bibfnamefont{K.~B.} \bibnamefont{Cooper}},
  \bibinfo{author}{\bibfnamefont{J.~P.} \bibnamefont{Eisenstein}},
  \bibinfo{author}{\bibfnamefont{L.~N.} \bibnamefont{Pfeiffer}},
  \bibnamefont{and} \bibinfo{author}{\bibfnamefont{K.~W.} \bibnamefont{West}},
  \emph{\bibinfo{title}{Evidence for an Anisotropic State of Two-Dimensional
  Electrons in High Landau Levels}}, \bibinfo{journal}{Phys. Rev. Lett.}
  \textbf{\bibinfo{volume}{82}}, \bibinfo{pages}{394} (\bibinfo{year}{1999}).

\bibitem[{\citenamefont{Du et~al.}(1999)\citenamefont{Du, Tsui, Stormer,
  Pfeiffer, Baldwin, and West}}]{du:1999}
\bibinfo{author}{\bibfnamefont{R.~R.} \bibnamefont{Du}},
  \bibinfo{author}{\bibfnamefont{D.~C.} \bibnamefont{Tsui}},
  \bibinfo{author}{\bibfnamefont{H.~L.} \bibnamefont{Stormer}},
  \bibinfo{author}{\bibfnamefont{L.~N.} \bibnamefont{Pfeiffer}},
  \bibinfo{author}{\bibfnamefont{K.~W.} \bibnamefont{Baldwin}},
  \bibnamefont{and} \bibinfo{author}{\bibfnamefont{K.~W.} \bibnamefont{West}},
  \emph{\bibinfo{title}{Strongly anisotropic transport in higher
  two-dimensional Landau levels}}, \bibinfo{journal}{Solid State Commun.}
  \textbf{\bibinfo{volume}{109}}, \bibinfo{pages}{389} (\bibinfo{year}{1999}).

\bibitem[{not({\natexlab{b}})}]{note:hb}
\bibinfo{note}{Signatures of anomalous nematic states have also been observed
  in Hall bar geometry.}

\bibitem[{\citenamefont{Kim et~al.}(2019)\citenamefont{Kim, Balram, Taniguchi,
  Watanabe, Jain, and Smet}}]{kim:2019}
\bibinfo{author}{\bibfnamefont{Y.}~\bibnamefont{Kim}},
  \bibinfo{author}{\bibfnamefont{A.~C.} \bibnamefont{Balram}},
  \bibinfo{author}{\bibfnamefont{T.}~\bibnamefont{Taniguchi}},
  \bibinfo{author}{\bibfnamefont{K.}~\bibnamefont{Watanabe}},
  \bibinfo{author}{\bibfnamefont{J.~K.} \bibnamefont{Jain}}, \bibnamefont{and}
  \bibinfo{author}{\bibfnamefont{J.~H.} \bibnamefont{Smet}},
  \emph{\bibinfo{title}{Even denominator fractional quantum Hall states in
  higher Landau levels of graphene}}, \bibinfo{journal}{Nat. Phys.}
  \textbf{\bibinfo{volume}{15}}, \bibinfo{pages}{154} (\bibinfo{year}{2019}).

\bibitem[{\citenamefont{Hossain et~al.}(2018)\citenamefont{Hossain, Ma, Chung,
  Pfeiffer, West, Baldwin, and Shayegan}}]{hossain:2019}
\bibinfo{author}{\bibfnamefont{M.~S.} \bibnamefont{Hossain}},
  \bibinfo{author}{\bibfnamefont{M.~K.} \bibnamefont{Ma}},
  \bibinfo{author}{\bibfnamefont{Y.~J.} \bibnamefont{Chung}},
  \bibinfo{author}{\bibfnamefont{L.~N.} \bibnamefont{Pfeiffer}},
  \bibinfo{author}{\bibfnamefont{K.~W.} \bibnamefont{West}},
  \bibinfo{author}{\bibfnamefont{K.~W.} \bibnamefont{Baldwin}},
  \bibnamefont{and} \bibinfo{author}{\bibfnamefont{M.}~\bibnamefont{Shayegan}},
  \emph{\bibinfo{title}{Unconventional Anisotropic Even-Denominator Fractional
  Quantum Hall State in a System with Mass Anisotropy}},
  \bibinfo{journal}{Phys. Rev. Lett.} \textbf{\bibinfo{volume}{121}},
  \bibinfo{pages}{256601} (\bibinfo{year}{2018}).

\bibitem[{\citenamefont{Xia et~al.}(2011)\citenamefont{Xia, Eisenstein,
  Pfeiffer, and West}}]{xia:2011}
\bibinfo{author}{\bibfnamefont{J.}~\bibnamefont{Xia}},
  \bibinfo{author}{\bibfnamefont{J.~P.} \bibnamefont{Eisenstein}},
  \bibinfo{author}{\bibfnamefont{L.~N.} \bibnamefont{Pfeiffer}},
  \bibnamefont{and} \bibinfo{author}{\bibfnamefont{K.~W.} \bibnamefont{West}},
  \emph{\bibinfo{title}{Evidence for a fractionally quantized Hall state with
  anisotropic longitudinal transport}}, \bibinfo{journal}{Nat. Phys.}
  \textbf{\bibinfo{volume}{7}}, \bibinfo{pages}{845} (\bibinfo{year}{2011}).

\bibitem[{\citenamefont{Liu et~al.}(2013)\citenamefont{Liu, Hasdemir, Shayegan,
  Pfeiffer, West, and Baldwin}}]{liu:2013b}
\bibinfo{author}{\bibfnamefont{Y.}~\bibnamefont{Liu}},
  \bibinfo{author}{\bibfnamefont{S.}~\bibnamefont{Hasdemir}},
  \bibinfo{author}{\bibfnamefont{M.}~\bibnamefont{Shayegan}},
  \bibinfo{author}{\bibfnamefont{L.~N.} \bibnamefont{Pfeiffer}},
  \bibinfo{author}{\bibfnamefont{K.~W.} \bibnamefont{West}}, \bibnamefont{and}
  \bibinfo{author}{\bibfnamefont{K.~W.} \bibnamefont{Baldwin}},
  \emph{\bibinfo{title}{Evidence for a 5/2 fractional quantum Hall nematic
  state in parallel magnetic fields}}, \bibinfo{journal}{Phys. Rev. B}
  \textbf{\bibinfo{volume}{88}}, \bibinfo{pages}{035307}
  (\bibinfo{year}{2013}).

\bibitem[{not({\natexlab{c}})}]{note:f}
\bibinfo{note}{Anomalous nematic states are very fragile and the $\ryy$
  maxima near $\nu = i/2$ are more elusive than the $\rxx$ minima. As for other
  fragile states in quantum Hall systems forming below 0.1 K, the uniformity of
  the carrier density is obviously an important factor. Another requirement is a good sample
  ``state'' which, we believe, is determined by the disorder landscape. The
  latter, in turn, sensitively depends on the details of both cooldown and
  illumination procedures, which are known to produce charge redistribution between the quantum weel, the doping layers, and the sample surface \citep{gamez:2013,samani:2014,fu:2018b}, thereby leading to
  different degrees of screening of the disorder potential \citep{sammon:2018}. 
  Nevertheless, after multiple cooldowns of different samples we are confident
  that the phenomenon is generic.}

\bibitem[{not({\natexlab{d}})}]{note:as}
\bibinfo{note}{We are aware of only two experiments which observed maximum
  resistance anisotropy at $\nus > 0.5$ \citep{shi:2017d,qian:2017}.}

\bibitem[{not({\natexlab{e}})}]{note:to}
\bibinfo{note}{A turnover in the temperature dependence has been observed
  previously \citep{cooper:thesis, qian:2017}, but no local resistance extrema near $\nus = 0.5$ have been reported to date.}

\bibitem[{\citenamefont{MacDonald and Fisher}(2000)}]{macdonald:2000}
\bibinfo{author}{\bibfnamefont{A.~H.} \bibnamefont{MacDonald}}
  \bibnamefont{and} \bibinfo{author}{\bibfnamefont{M.~P.~A.}
  \bibnamefont{Fisher}}, \emph{\bibinfo{title}{Quantum theory of quantum Hall
  smectics}}, \bibinfo{journal}{Phys. Rev. B} \textbf{\bibinfo{volume}{61}},
  \bibinfo{pages}{5724} (\bibinfo{year}{2000}).

\bibitem[{not({\natexlab{f}})}]{note:mf}
\bibinfo{note}{At $\nus = 1/2$, \rref{macdonald:2000} predicts $\rho_{xx}
  \propto T^{-\alpha}$ and $\rho_{yy} \propto T^{\alpha}$ with $\alpha \approx
  0.5$.}

\bibitem[{\citenamefont{Qian et~al.}(2017)\citenamefont{Qian, Nakamura,
  Fallahi, Gardner, and Manfra}}]{qian:2017}
\bibinfo{author}{\bibfnamefont{Q.}~\bibnamefont{Qian}},
  \bibinfo{author}{\bibfnamefont{J.}~\bibnamefont{Nakamura}},
  \bibinfo{author}{\bibfnamefont{S.}~\bibnamefont{Fallahi}},
  \bibinfo{author}{\bibfnamefont{G.~C.} \bibnamefont{Gardner}},
  \bibnamefont{and} \bibinfo{author}{\bibfnamefont{M.~J.}
  \bibnamefont{Manfra}}, \emph{\bibinfo{title}{Possible nematic to smectic
  phase transition in a two-dimensional electron gas at half-filling}},
  \bibinfo{journal}{Nat. Commun.} \textbf{\bibinfo{volume}{8}},
  \bibinfo{pages}{1536} (\bibinfo{year}{2017}).

\bibitem[{\citenamefont{Shi et~al.}(2017{\natexlab{a}})\citenamefont{Shi,
  Zudov, Qian, Watson, and Manfra}}]{shi:2017c}
\bibinfo{author}{\bibfnamefont{Q.}~\bibnamefont{Shi}},
  \bibinfo{author}{\bibfnamefont{M.~A.} \bibnamefont{Zudov}},
  \bibinfo{author}{\bibfnamefont{Q.}~\bibnamefont{Qian}},
  \bibinfo{author}{\bibfnamefont{J.~D.} \bibnamefont{Watson}},
  \bibnamefont{and} \bibinfo{author}{\bibfnamefont{M.~J.}
  \bibnamefont{Manfra}}, \emph{\bibinfo{title}{Effect of density on quantum
  Hall stripe orientation in tilted magnetic fields}}, \bibinfo{journal}{Phys.
  Rev. B} \textbf{\bibinfo{volume}{95}}, \bibinfo{pages}{161303(R)}
  (\bibinfo{year}{2017}{\natexlab{a}}).

\bibitem[{\citenamefont{Fu et~al.}(2019)\citenamefont{Fu, Shi, Zudov, Gardner,
  Watson, and Manfra}}]{fu:2019a}
\bibinfo{author}{\bibfnamefont{X.}~\bibnamefont{Fu}},
  \bibinfo{author}{\bibfnamefont{Q.}~\bibnamefont{Shi}},
  \bibinfo{author}{\bibfnamefont{M.~A.} \bibnamefont{Zudov}},
  \bibinfo{author}{\bibfnamefont{G.~C.} \bibnamefont{Gardner}},
  \bibinfo{author}{\bibfnamefont{J.~D.} \bibnamefont{Watson}},
  \bibnamefont{and} \bibinfo{author}{\bibfnamefont{M.~J.}
  \bibnamefont{Manfra}}, \emph{\bibinfo{title}{Two- and three-electron bubbles
  in
  ${\mathrm{Al}}_{x}{\mathrm{Ga}}_{1\ensuremath{-}x}\mathrm{As}$/${\mathrm{Al}}_{0.24}{\mathrm{Ga}}_{0.76}\mathrm{As}$
  quantum wells}}, \bibinfo{journal}{Phys. Rev. B}
  \textbf{\bibinfo{volume}{99}}, \bibinfo{pages}{161402}
  (\bibinfo{year}{2019}).

\bibitem[{\citenamefont{Shi et~al.}(2016{\natexlab{a}})\citenamefont{Shi,
  Zudov, Watson, Gardner, and Manfra}}]{shi:2016b}
\bibinfo{author}{\bibfnamefont{Q.}~\bibnamefont{Shi}},
  \bibinfo{author}{\bibfnamefont{M.~A.} \bibnamefont{Zudov}},
  \bibinfo{author}{\bibfnamefont{J.~D.} \bibnamefont{Watson}},
  \bibinfo{author}{\bibfnamefont{G.~C.} \bibnamefont{Gardner}},
  \bibnamefont{and} \bibinfo{author}{\bibfnamefont{M.~J.}
  \bibnamefont{Manfra}}, \emph{\bibinfo{title}{Reorientation of quantum Hall
  stripes within a partially filled Landau level}}, \bibinfo{journal}{Phys.
  Rev. B} \textbf{\bibinfo{volume}{93}}, \bibinfo{pages}{121404(R)}
  (\bibinfo{year}{2016}{\natexlab{a}}).

\bibitem[{\citenamefont{Shi et~al.}(2016{\natexlab{b}})\citenamefont{Shi,
  Zudov, Watson, Gardner, and Manfra}}]{shi:2016c}
\bibinfo{author}{\bibfnamefont{Q.}~\bibnamefont{Shi}},
  \bibinfo{author}{\bibfnamefont{M.~A.} \bibnamefont{Zudov}},
  \bibinfo{author}{\bibfnamefont{J.~D.} \bibnamefont{Watson}},
  \bibinfo{author}{\bibfnamefont{G.~C.} \bibnamefont{Gardner}},
  \bibnamefont{and} \bibinfo{author}{\bibfnamefont{M.~J.}
  \bibnamefont{Manfra}}, \emph{\bibinfo{title}{Evidence for a new symmetry
  breaking mechanism reorienting quantum Hall nematics}},
  \bibinfo{journal}{Phys. Rev. B} \textbf{\bibinfo{volume}{93}},
  \bibinfo{pages}{121411(R)} (\bibinfo{year}{2016}{\natexlab{b}}).

\bibitem[{\citenamefont{Fradkin and Kivelson}(1999)}]{fradkin:1999}
\bibinfo{author}{\bibfnamefont{E.}~\bibnamefont{Fradkin}} \bibnamefont{and}
  \bibinfo{author}{\bibfnamefont{S.~A.} \bibnamefont{Kivelson}},
  \emph{\bibinfo{title}{Liquid-crystal phases of quantum Hall systems}},
  \bibinfo{journal}{Phys. Rev. B} \textbf{\bibinfo{volume}{59}},
  \bibinfo{pages}{8065} (\bibinfo{year}{1999}).

\bibitem[{\citenamefont{Gamez and Muraki}(2013)}]{gamez:2013}
\bibinfo{author}{\bibfnamefont{G.}~\bibnamefont{Gamez}} \bibnamefont{and}
  \bibinfo{author}{\bibfnamefont{K.}~\bibnamefont{Muraki}},
  \emph{\bibinfo{title}{$\ensuremath{\nu}=5/2$ fractional quantum Hall state in
  low-mobility electron systems: Different roles of disorder}},
  \bibinfo{journal}{Phys. Rev. B} \textbf{\bibinfo{volume}{88}},
  \bibinfo{pages}{075308} (\bibinfo{year}{2013}).

\bibitem[{\citenamefont{Samani et~al.}(2014)\citenamefont{Samani, Rossokhaty,
  Sajadi, L\"uscher, Folk, Watson, Gardner, and Manfra}}]{samani:2014}
\bibinfo{author}{\bibfnamefont{M.}~\bibnamefont{Samani}},
  \bibinfo{author}{\bibfnamefont{A.~V.} \bibnamefont{Rossokhaty}},
  \bibinfo{author}{\bibfnamefont{E.}~\bibnamefont{Sajadi}},
  \bibinfo{author}{\bibfnamefont{S.}~\bibnamefont{L\"uscher}},
  \bibinfo{author}{\bibfnamefont{J.~A.} \bibnamefont{Folk}},
  \bibinfo{author}{\bibfnamefont{J.~D.} \bibnamefont{Watson}},
  \bibinfo{author}{\bibfnamefont{G.~C.} \bibnamefont{Gardner}},
  \bibnamefont{and} \bibinfo{author}{\bibfnamefont{M.~J.}
  \bibnamefont{Manfra}}, \emph{\bibinfo{title}{Low-temperature illumination and
  annealing of ultrahigh quality quantum wells}}, \bibinfo{journal}{Phys. Rev.
  B} \textbf{\bibinfo{volume}{90}}, \bibinfo{pages}{121405}
  (\bibinfo{year}{2014}).

\bibitem[{\citenamefont{Fu et~al.}(2018)\citenamefont{Fu, Riedl, Borisov,
  Zudov, Watson, Gardner, Manfra, Baldwin, Pfeiffer, and West}}]{fu:2018b}
\bibinfo{author}{\bibfnamefont{X.}~\bibnamefont{Fu}},
  \bibinfo{author}{\bibfnamefont{A.}~\bibnamefont{Riedl}},
  \bibinfo{author}{\bibfnamefont{M.}~\bibnamefont{Borisov}},
  \bibinfo{author}{\bibfnamefont{M.~A.} \bibnamefont{Zudov}},
  \bibinfo{author}{\bibfnamefont{J.~D.} \bibnamefont{Watson}},
  \bibinfo{author}{\bibfnamefont{G.}~\bibnamefont{Gardner}},
  \bibinfo{author}{\bibfnamefont{M.~J.} \bibnamefont{Manfra}},
  \bibinfo{author}{\bibfnamefont{K.~W.} \bibnamefont{Baldwin}},
  \bibinfo{author}{\bibfnamefont{L.~N.} \bibnamefont{Pfeiffer}},
  \bibnamefont{and} \bibinfo{author}{\bibfnamefont{K.~W.} \bibnamefont{West}},
  \emph{\bibinfo{title}{Effect of illumination on quantum lifetime in GaAs
  quantum wells}}, \bibinfo{journal}{Phys. Rev. B}
  \textbf{\bibinfo{volume}{98}}, \bibinfo{pages}{195403}
  (\bibinfo{year}{2018}).

\bibitem[{\citenamefont{Sammon et~al.}(2018)\citenamefont{Sammon, Zudov, and
  Shklovskii}}]{sammon:2018}
\bibinfo{author}{\bibfnamefont{M.}~\bibnamefont{Sammon}},
  \bibinfo{author}{\bibfnamefont{M.~A.} \bibnamefont{Zudov}}, \bibnamefont{and}
  \bibinfo{author}{\bibfnamefont{B.~I.} \bibnamefont{Shklovskii}},
  \emph{\bibinfo{title}{Mobility and quantum mobility of modern GaAs/AlGaAs
  heterostructures}}, \bibinfo{journal}{Phys. Rev. Materials}
  \textbf{\bibinfo{volume}{2}}, \bibinfo{pages}{064604} (\bibinfo{year}{2018}).

\bibitem[{\citenamefont{Shi et~al.}(2017{\natexlab{b}})\citenamefont{Shi,
  Zudov, Friess, Smet, Watson, Gardner, and Manfra}}]{shi:2017d}
\bibinfo{author}{\bibfnamefont{Q.}~\bibnamefont{Shi}},
  \bibinfo{author}{\bibfnamefont{M.~A.} \bibnamefont{Zudov}},
  \bibinfo{author}{\bibfnamefont{B.}~\bibnamefont{Friess}},
  \bibinfo{author}{\bibfnamefont{J.}~\bibnamefont{Smet}},
  \bibinfo{author}{\bibfnamefont{J.~D.} \bibnamefont{Watson}},
  \bibinfo{author}{\bibfnamefont{G.~C.} \bibnamefont{Gardner}},
  \bibnamefont{and} \bibinfo{author}{\bibfnamefont{M.~J.}
  \bibnamefont{Manfra}}, \emph{\bibinfo{title}{Apparent temperature-induced
  reorientation of quantum Hall stripes}}, \bibinfo{journal}{Phys. Rev. B}
  \textbf{\bibinfo{volume}{95}}, \bibinfo{pages}{161404(R)}
  (\bibinfo{year}{2017}{\natexlab{b}}).

\bibitem[{\citenamefont{Cooper}(2003)}]{cooper:thesis}
\bibinfo{author}{\bibfnamefont{K.}~\bibnamefont{Cooper}},
  \emph{\bibinfo{title}{New Phases of Two-Dimensional Electrons in Excited
  Landau Levels}}, Ph.D. thesis, \bibinfo{school}{California Institute of
  Technology} (\bibinfo{year}{2003}).

\end{thebibliography}

\end{document}